\documentclass[manuscript]{aastex}

\slugcomment{}

\shorttitle{Deep galaxy LF}
\shortauthors{Harsono \& De Propris}

\begin{document}


\title{The luminosity function of galaxies to $M_{BgVriz} \sim -14$ in
$z \sim 0.3$ clusters}


\author{Daniel Harsono\altaffilmark{1}}
\affil{Department of Physics and Astronomy, UCLA, 430 Portola Plaza, Box 951547, Los Angeles, 
CA 90095-1547, USA}
\email{dharsono@ucla.edu}

\and

\author{Roberto De Propris}
\affil{Cerro Tololo Inter-American Observatory, Casilla 603, La Serena, Chile}
\email{rdepropris@ctio.noao.edu}


\altaffiltext{1}{Current address: Leiden Observatory, Leiden University, Postbus 9513, 
NL-2300 RA Leiden, The Netherlands.}


\begin{abstract}

We present deep composite luminosity functions in $B$, $g$, $V$, $r$, $i$
and $z$ for six clusters at $0.14 < z < 0.40$ observed with the Hubble Space
Telescope Advanced Camera for Surveys. The luminosity functions reach to
absolute magnitude of $\sim -14 + 5\log h$ mag. and are well fitted by a single
Schechter function with $M^*_{BgVriz}=-19.8,\,-20.9\,-21.9,\,-22.0,\,-21.7,\,
-22.3$ mag. and $\alpha \sim -1.3$ (in all bands). The observations suggest that
the galaxy luminosity function is dominated by objects on the red sequence
to at least 6 mags. below the $L^*$ point. Comparison with local data shows that
the red sequence is well established at least at $z \sim 0.3$ down to $\sim 1/600^{th}$
of the luminosity of the Milky Way and that galaxies down to the regime of dwarf
spheroidals have been completely assembled in clusters at this redshift.
We do not detect a steepening of the luminosity function at $M > -16$ as
is observed locally. If the faint end upturn is real, the steepening of
the luminosity function must be due to a newly infalling population of
faint dwarf galaxies.

\end{abstract}





\section{Introduction}

The galaxy luminosity function (hereafter LF) provides a $0^{th}$ order
description of the gross properties of galaxy populations. Although it 
may be regarded as somewhat simplistic, in reducing galaxy properties 
to the two numbers ($M^*$ and $\alpha$) that describe the LF in the
\cite{schechter76} form, reproducing the observed LF is a fundamental
test for any viable theory of galaxy formation, and one that has
proven surprisingly complex to achieve until relatively recently (e.g.,
\citealt{croton06}).

Clusters of galaxies are, in several ways, ideal laboratories to study 
the LF and its evolution. Cluster galaxies may be considered as a volume
limited sample of objects observed at the same cosmic epoch and occupy
a relatively `constant' environment corresponding to the highest density
peaks at the redshift of observation. Operationally, cluster members have
a much higher surface density on the sky than the surrounding field and
often exhibit distinctive morphologies and colors (see Fig.~1 below). It
is therefore possible to establish membership in a cluster (at least in a
statistical sense) without resorting to observationally expensive redshift
surveys.

On the other hand, clusters are obviously special places, containing only a 
small percentage ($5$--$10\%$) of all galaxies in the Universe. Their galaxy 
populations clearly implicate the cluster environment in a variety of processes 
that affect the morphological evolution of galaxies (e.g., \citealt{dressler80})
and suppress or modify their star formation history \citep{lewis02,gomez03}.
In order to use clusters as probes of galaxy evolution it is therefore
important to understand the effects of environment and carry out careful
comparisons with field samples. For local samples, this is now possible
thanks to large redshift surveys such as 2dF \citep{colless01} and SDSS
\citep{york00}.

Although bright galaxies appear to have formed the majority of their stellar
populations and assembled their mass at least by $z=1.5$ (e.g., \citealt{andreon06a,
depropris07,muzzin08} and references therein), there is considerable evidence
that fainter galaxies undergo significant evolution since $z=1$. The red sequence
in clusters in the EDisCS sample appears to be truncated at faint magnitudes
\citep{delucia07}, while \cite{stott07} show that the luminosity function of
red cluster galaxies flattens at higher redshifts, as do \cite{krick08} for
clusters in a deep {\it Spitzer} field.  On the other hand, there are clusters 
with well-formed red sequences at redshifts approaching $1$ as well \citep{andreon06b,
crawford08}. It may not be surprising that the faint end of the LF shows considerable
variations from cluster to cluster, as one expects that dwarfs are more sensitive
to environmental effects. Although the bright end of the LF ($M < -16$) is broadly
universal, the faint end varies between clusters and may even change radially within
clusters \citep{popesso06,barkhouse07}.

The evolution of the faint end slope of the total LF provides an interesting test
of galaxy formation models. Even if the red sequence is truncated at faint luminosities
at earlier epochs, one expects a steepening faint end as established by the initial
power spectrum of fluctuations, which is very steep \citep{khochfar07}. In higher
redshift clusters, there should be a steeper LF and a larger fraction of bluer
dwarfs.

The purpose of this paper is to explore the evolution of the LF to very faint
magnitudes in a sample of $z \sim 0.3$ clusters observed with the Hubble Space
Telescope (HST), reaching well below the $L^*$ point and into the domain of
dwarf spheroidal galaxies, using a multicolor imaging sample. This will allow
to test scenarios for the evolution of dwarf galaxy populations in $\Lambda$CDM
hierarchical models. The structure of this paper is as follows: we describe
the data and photometry in the next section. We derive the composite LFs and
then discuss the results in the context of galaxy formation models. We adopt 
the `consensus' cosmology with $\Omega_M=0.27$, $\Omega_{\Lambda}=0.73$ and
use H$_0$=100 km s$^{-1}$ Mpc$^{-1}$. All data are photometrically calibrated
to the AB system, using the latest ACS zeropoints published on the HST Instrument
Web page.

\section{Observations and Data Analysis}

Observations for this paper consist of deep HST imaging of six clusters at
$0.14 < z < 0.40$ taken with the Advanced Camera for Surveys (ACS) in the
$B$ (F435W), $g$ (F475W), $V$ (F555W), $r$ (F625W), $i$ (F775W) and $z$
(F850LP) bands with exposure times of 5,000--10,000 seconds in each band.
Each cluster was observed in a single ACS exposure, covering about $200''$
on the sky. The data were originally taken for studies of gravitational lenses 
in these clusters. Table 1 shows the clusters observed, their redshifts, exposure
times in each band, and HST proposal IDs. Table 2 summarizes some essential
physical characteristics for the observed clusters. Most data are from the
compilation of \cite{wu99}, from which $r_{200}$ can be estimated via the
formula of \cite{carlberg97}, except for A1413 where $r_{200}$ was derived by
\cite{pointecouteau05} from the X-ray profile. We were unable to find data in
the literature for A1703.

The data were retrieved from the HST archive as fully reduced and flatfielded
files (*.flt) and then processed through the {\tt Multidrizzle} algorithm
\citep{koekemoer02} to produce fully registered and co-added images for each
band. Fig.~1 shows color composites for our data. Full resolution JPEGs will
be made available on the $AJ$ web site.

Detection of objects and photometry were carried out using Sextractor \citep{
bertin96}. We experimented with Sextractor parameters to maximize our detections
and minimize the number of spurious objects. All detections were visually 
examined to eliminate noise spikes, bleed trails from bright stars and 
other contaminants, especially the numerous arclets present in the images. 
The Sextractor parameters eventually employed are shown in Table 3. We used 
the same parameters for all images.
 
For each object we compute a total (Kron-like) magnitude and an aperture
magnitude of area equivalent to the minimum number of connected pixels needed
for detection. This provides an estimate of the central surface brightness
for each object. The motivation behind this procedure is as follows. Detections 
of objects in deep images depends on their total magnitude but also on their 
central surface brightness. An object could be brighter than the magnitude 
threshold but be lost in the night sky because of low surface brightness (see 
discussion in \citealt{cross02}). For this reason we need to determine both a 
limiting total magnitude for completeness and a surface brightness threshold 
for detection.

We do this by plotting central surface brightness vs. total magnitude for
all bands in Fig.~2. To save space we only show the data for Abell 1703, but
equivalent figures for all other clusters are available on the $AJ$ web site.
In these figures the sequence at high surface brightness that provides an upper
limit to the scatter plot is caused by stars. We can therefore discriminate 
against stellar contamination in this way (cf. \citealt{garilli99}). The surface 
brightness threshold is chosen empirically. The limit is selected 
to include as many objects as possible, but avoiding regions of the parameter 
space where the sample is obviously very incomplete. Selection lines in surface 
brightness, apparent magnitude and the star-galaxy separation line are shown in 
Fig.~2.

In order to determine a limiting magnitude, we plot the raw number counts for
galaxies in the Abell 1703 field in Fig.~3. As for Fig.~2, figures for all the
other clusters are made available electronically. The completeness magnitude is 
chosen to lie about 0.5 mag.~brighter than the luminosity at which counts start 
to decrease, in order to select a highly complete sample of objects.

The only way to establish cluster membership for faint galaxies is by statistical
background subtraction. In order to do so, we need to analyze `blank' (cluster-less)
comparison fields of similar or greater photometric depth. We choose to use the
two GOODS fields \citep{giavalisco04} as these are the deepest and widest fields
where HST multicolor photometry is available.

We used the same Sextractor parameters as we used for the cluster fields and
impose the same selection limits on the GOODS fields as we did for the target
fields. Fig.~4 shows the detection and selection plots for the GOODS fields
(similar to Fig.~2 above). The fields are much deeper than the cluster fields
we use. It must be noted that we plot only $10\%$ of the detections in the GOODS
fields in Fig.~4 to avoid saturating the figure.

Fig.~5 plots galaxy number counts for the GOODS fields and comparison galaxy
number counts from the literature. The HST data were corrected to the SDSS
and Johnson-Cousins systems using the transformations tabulated in \cite{holberg06}.
GOODS number counts were fitted with a quadratic of the form $a_0+a_1 x+a_2x^2$.
The fit was carried out over galaxies brighter than the completeness limit shown
in Fig.~4 and fainter than m$\sim 21$ mag. as GOODS counts for brighter galaxies
are affected by small scale clustering. Table 4 shows the values of the coefficients 
for the quadratic fits.

\section{Luminosity Functions}

We now subtract the scaled contribution from fore/background counts from 
galaxy counts. Errors in galaxy counts for clusters and the GOODS fields
are assumed to be Poissonian. We used our smoothed quadratic for the background
counts, extrapolating for galaxies brighter than $m \sim 21$ mag. Field counts
at these bright limits and over the small ($\sim 9$ arcmin$^2$) covered by 
ACS are very small in any case. Clustering errors over the cluster and GOODS
fields are estimated following the counts-in-cells approach and the fitting
formula described by \cite{huang97}. 

It may be possible to further refine the selection of cluster members using
photometric redshifts or color cuts. Contamination of the LF by background
galaxies has been suggested as a possible cause of the faint end upturn (see
below) by \cite{valotto01}. However, this would lead to loss of generality.
The argument by \cite{valotto01} may be correct for nearby clusters (although
\citealt{andreon05} point out that the simulations used by \citealt{valotto01}
are unrealistic and that the background subtraction approach used here works
as long as proper care is taken of statistical uncertainties, see their \S 3.3.6),
but it is not appropriate for these distant objects where the redshift is much
larger than the typical scale of structure in the Universe (so that the counts
should be smooth).

There are some mismatches between the GOODS data (used for background subtraction)
and the cluster data. For the $g$ (F435W) and $r$ (F625W) bands, we can use the
GOODS F475W and F606W counts (respectively), which are close matches to the bands
used, however, for F555W ($V$) we used the average of the F475W and F606W counts.
The LF in this bands is more uncertain, because we do not have an appropriate
background field.

We $k$ and $e$ correct the LFs to $z=0$ for ease of comparison with local
data. In order to do so we use a \cite{bruzual03} model for a solar metallicity
elliptical formed at $z=3$ with an e-folding time of 1 Gyr. This is appropriate
to the red ellipticals that dominate the bright end of cluster LFs. In
effect this approach allows us to consider the differential evolution,
if any, between the dwarfs and the giants, who appear to consist mainly
of old galaxies with little or no recent star formation.

The LF of each individual cluster is relatively uncertain. We therefore
build composite LFs in each band following the method by \cite{colless89}
in order to average out errors. The method assumes that the LF is broadly
universal, but this appears to be borne out by observations at low redshift,
at least for relatively bright galaxies \citep{popesso06,barkhouse07}. 

Composite LFs are built as follows: the number of cluster galaxies in the
$j^{th}$ magnitude bin of the composite LF is given by:

\begin{equation}
N_{cj}={N_{c0} \over m_j} {\sum_i {N_{ij} \over N_{i0}}}
\end{equation}

where $N_{ij}$ is the (background corrected) number of galaxies in the $j^{th}$ 
magnitude bin of the LF of the $i^{th}$ cluster LF, $N_{i0}$ is the normalization
of the cluster LF (to take care of the different richnesses of the clusters) and
is taken to be the background corrected number of cluster galaxies brighter than
$M_B=-19$ (or its equivalent in the other bands, assuming the color of an old
elliptical), $m_j$ is the number of $i$ clusters contributing to the $j^{th}$ 
magnitude bin of the composite LF and $N_{c0}$ is the sum of all the $i$ normalizations.

\begin{equation}
N_{c0}=\sum_i N_{i0}
\end{equation}

Essentially, this carried out a weighted average of the individual cluster LFs, scaled
by the number of galaxies (in each cluster) brighter than approximately the $L^*$ point.
The formal errors on $N_{cj}$ are computed as:

\begin{equation}
\delta N_{cj} = {N_{c0} \over m_j} \left[\sum_i \left({\delta N_{ij} \over N_{i0}} \right)^2  
\right]^{1/2}
\end{equation}

where $\delta N_{ij}$ are the errors on the number counts (including Poissonian and clustering
contributions) for the $j^{th}$ magnitude bin of the $i^{th}$ cluster LF.

For each cluster we build the composite LF by summing up the individual LFs in absolute
magnitude bins, after correcting for the distance modulus, extinction and $k+e$ correction
to the common redshift of $z=0$.

The resulting LFs are presented in Fig.~6, together with the 1, 2 and 3$\sigma$ error countours.
We fit the data to a single Schechter function (see discussion below, regarding the existence
of a faint end upturn). In some cases we exclude the fainter and/or brighter magnitude bin from
the fit, because these points are dominated by either the poorly sampled very massive galaxies or
the fainter objects that suffer from complex incompleteness due to magnitude and surface brightness
selection. The fit parameters are tabulated in Table 5, together with the $1\sigma$
errors for each parameter, marginalized over the remaining parameters.

\section{Discussion}

There are only a few deep cluster LFs having comparable depth to our own.
In general, we are in good agreement with previous work, based on smaller
and shallower samples \citep{driver94,trentham98a,boyce01,mercurio03,
pracy04,andreon05,harsono07}. The $M^*$ point we measure gets steadily 
brighter with increasing wavelength, tracking the color of old, passively 
evolving stellar populations that are the main component of bright
elliptical galaxies \citep{cool06}. The value we find is also consistent
with the local SDSS value by \cite{popesso06}: this is not surprising, as
it is well known that massive galaxies evolve passively since high redshift.

The main interest of this investigation is in the evolution of the LF slope.
The LF is well fitted by a single Schechter function, with a slope $\alpha
\sim -1.3$ which is essentially the same in all bands. This suggests that
the cluster LF is dominated by galaxies on the red sequence to a luminosity
of $M \sim -14$ in all bands, and that therefore the red sequence continues
at least to 6 magnitudes below the $L^*$ point even at $z \sim 0.3$. This
can be clearly seen in Fig.~7 where we show the $g-r$ vs. $z$ color-magnitude
relation for Abell 1703. The relation can be clearly followed at least to
$z=25$ with little scatter. A more detailed analysis of the color-magnitude
relation is deferred to a future paper (Harsono \& De Propris 2008, in preparation).
The red sequence appears to be well established and contain the majority
of cluster populations even at a lookback time of $\sim 3.5$ Gyrs (cf. \citealt{
andreon06c,eisenhardt07} for a similarly deep relation in local clusters).

The slope is very similar to the local value for $M < -16$ derived by
\cite{popesso06} and \cite{barkhouse07}. This argues that even galaxies
to $M \sim -14$ have formed their stellar populations and assembled the
majority of their mass at least at $z \sim 0.3$. We can therefore place
a significant lower limit to the assembly epoch of galaxies down to 1/600$^{th}$
of the mass of the Milky Way. Given the evidence for truncation of the red
sequence at $z \sim 0.8$ \citep{delucia07,stott07,krick08}, our data point
to the $0.3 < z < 0.8$ interval as a crucial epoch to investigate the formation of
the fainter galaxy populations.

The other issue we wish to address is the faint end upturn. This has been a
controversial subject, ever since its original discovery by \cite{driver94}
and \cite{depropris95}. There have been a series of claims and counterclaims
regarding the faint end slope and the existence of an upward inflection,
sometimes even in the same cluster. For instance, \cite{depropris98,
trentham98b,adami07,jenkins07} and \cite{milne07} observe a steep LF
at the faint end for the Coma cluster, while \cite{bernstein95,adami98}
and \cite{beijersbergen02} quote much flatter slopes. In Virgo, \cite{sabatini05}
claim a slope as steep as $\sim -1.6$, while \cite{rines08} derive $\alpha=-1.3$.
\cite{baldry08} review the existence of the upturn and the shape of the faint
end of the LF and conclude that there is a steep mass function. The observation
of the upturn in two composite LFs, derived by two different groups \citep{popesso06,
barkhouse07}, suggests that the upward inflection of the LF is real (however,
cf., \citealt{hilker03,penny08}).

Unlike \cite{popesso06} and \cite{barkhouse07} we do not find a steepening
of the LF at $M > -16$. However, we must consider that the LF upturn is more
pronounced at large clustercentric radii, while our fields generally cover 
only the central region of each cluster. The size of the ACS fields used
in this study covers between 350 and 750 $h^{-1}$ kpc on the side. In terms
of $R_{200}$ (the radius at which \citealt{popesso06} and \cite{barkhouse07}
normalize their LFs), the areas imaged by ACS cover between 20\% and 40\% of
the area of the cluster out to $r_{200}$. We therefore only derive LFs for
the cluster cores. This should not affect our comparisons with the bright
end of the LF ($M < -16$), as this does not seem to vary significantly 
with radius. However, the steep upturn claimed by \cite{popesso06} and
\cite{barkhouse07} is much more pronounced at large clustercentric
radii, i.e., in the cluster outskirts. On the other hand it is possible
to see the upturn even in the more central regions of the sample studied
by \cite{popesso06}, inside $\sim 0.3r_{200}$ (see Fig.~10 of \citealt{
popesso06}), so our data should exhibit an upturn at faint magnitudes,
although of course more clusters would be welcome to bolster the argument.

As we do not detect a steepening in the LF, we suggest that, if the
steepening is real, the faint galaxies contributing to the upturn 
consist of a population of recently infalling objects (e.g., \citealt{
smith08}) whose star formation is curtailed by the cluster environment.
This would be consistent with the observation that about 1/2 of the fainter
cluster dwarfs in Virgo and elsewhere have been forming stars until
relatively recently \citep{conselice01,conselice03}, until their star
formation stopped and their colors moved to the red sequence. The newly
infalling population may be the one now contributing to the steep upturn.
In a `downsizing' model, it may be expected that objects undergoing 
star formation suppression and cluster infall at the present epoch
would indeed tend to be among the fainter and less massive galaxies
and to have the steep LF characteristic of `pristine' CDM power spectra.

Our observations imply that the passive evolution of bright galaxies can
be extended to faint dwarfs, at least to $z \sim 0.3$ and suggest that 
the majority of galaxy evolution may have taken place at surprisingly early
epochs even for the least massive objects.





\acknowledgments

We would like to thank the referee for a very helpful report, that has considerably improved the
paper. We would also like to thank Stefano Andreon for having commented on a version of this paper.

All of the data presented in this paper were obtained from the Multimission Archive at the Space Telescope Science Institute (MAST). STScI is operated by the Association of Universities for Research in Astronomy, Inc., under NASA contract NAS5-26555. Support for MAST for non-HST data is provided by the NASA Office of Space Science via grant NAG5-7584 and by other grants and contracts.

We acknowledge use of the Cosmology Calculator \citep{wright06}.



{\it Facilities:} \facility{HST (ACS)}.

\clearpage

\begin{figure*}
\begin{tabular}{cc}
\end{tabular}
\caption{Colour images of target clusters. From left to right: A1413, A1689,
A2218, A1703, MS1358 and Cl0024. {\sc These figures exceed arXiv size limits
and cannot be compressed further. They will appear as JPEG in AJ. Please email
{\tt rdepropris@ctio.noao.edu} for the figures to be sent to you via FTP (provide
site) or email}}
\end{figure*}

\clearpage

\begin{figure*}
\begin{tabular}{cc}
\includegraphics[width=6cm,angle=270]{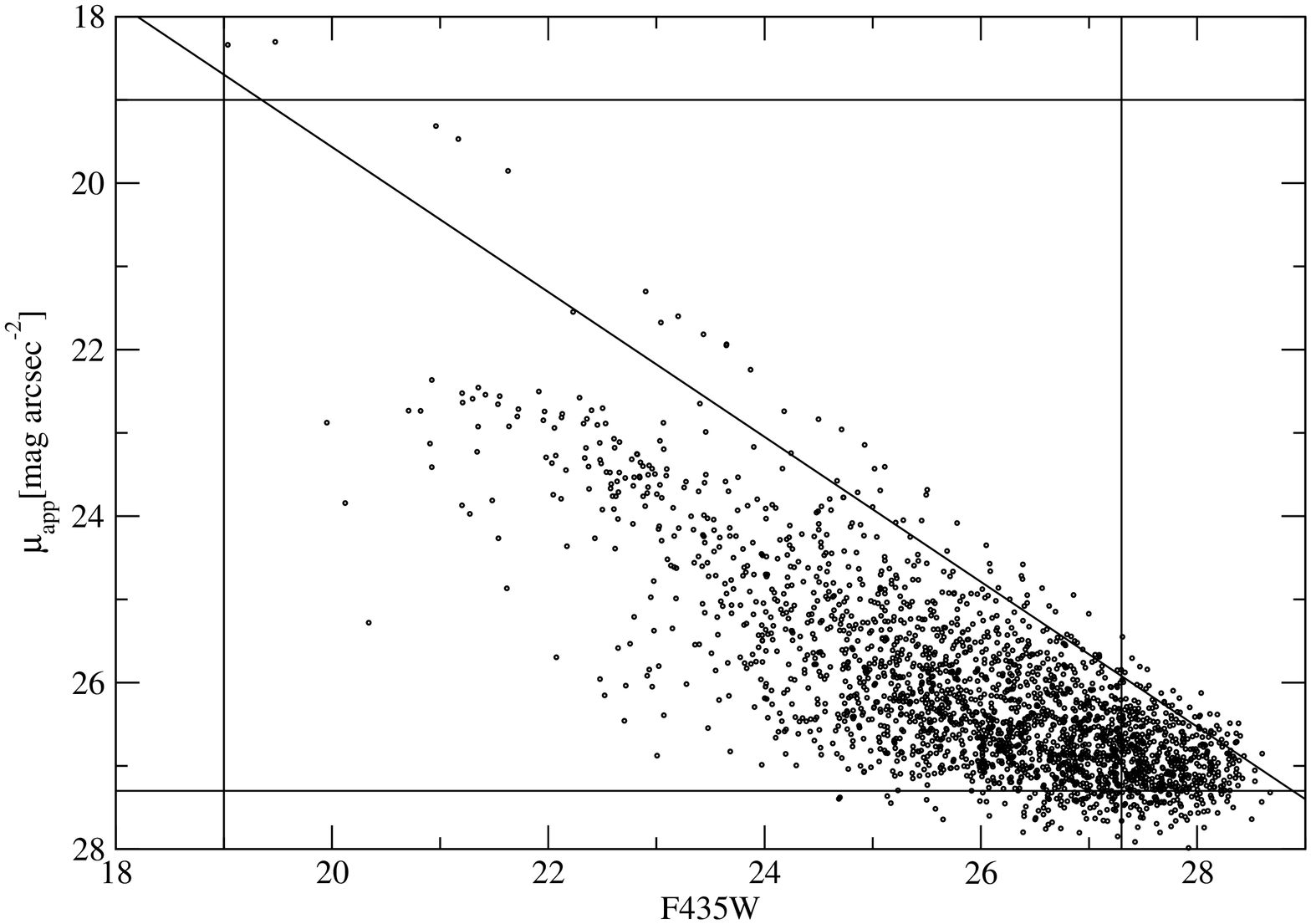} & \includegraphics[width=6cm,angle=270]{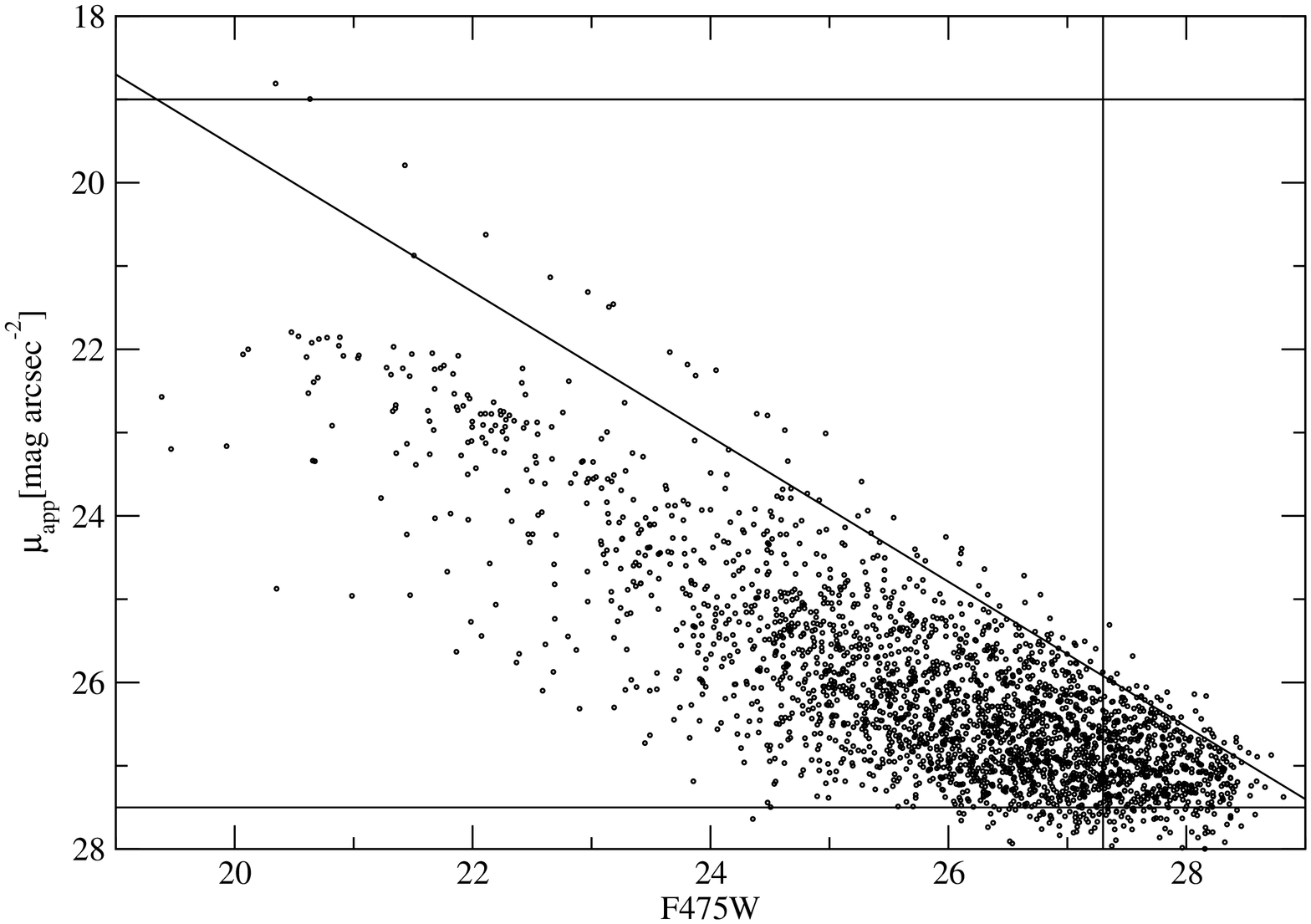} \\
\includegraphics[width=6cm,angle=270]{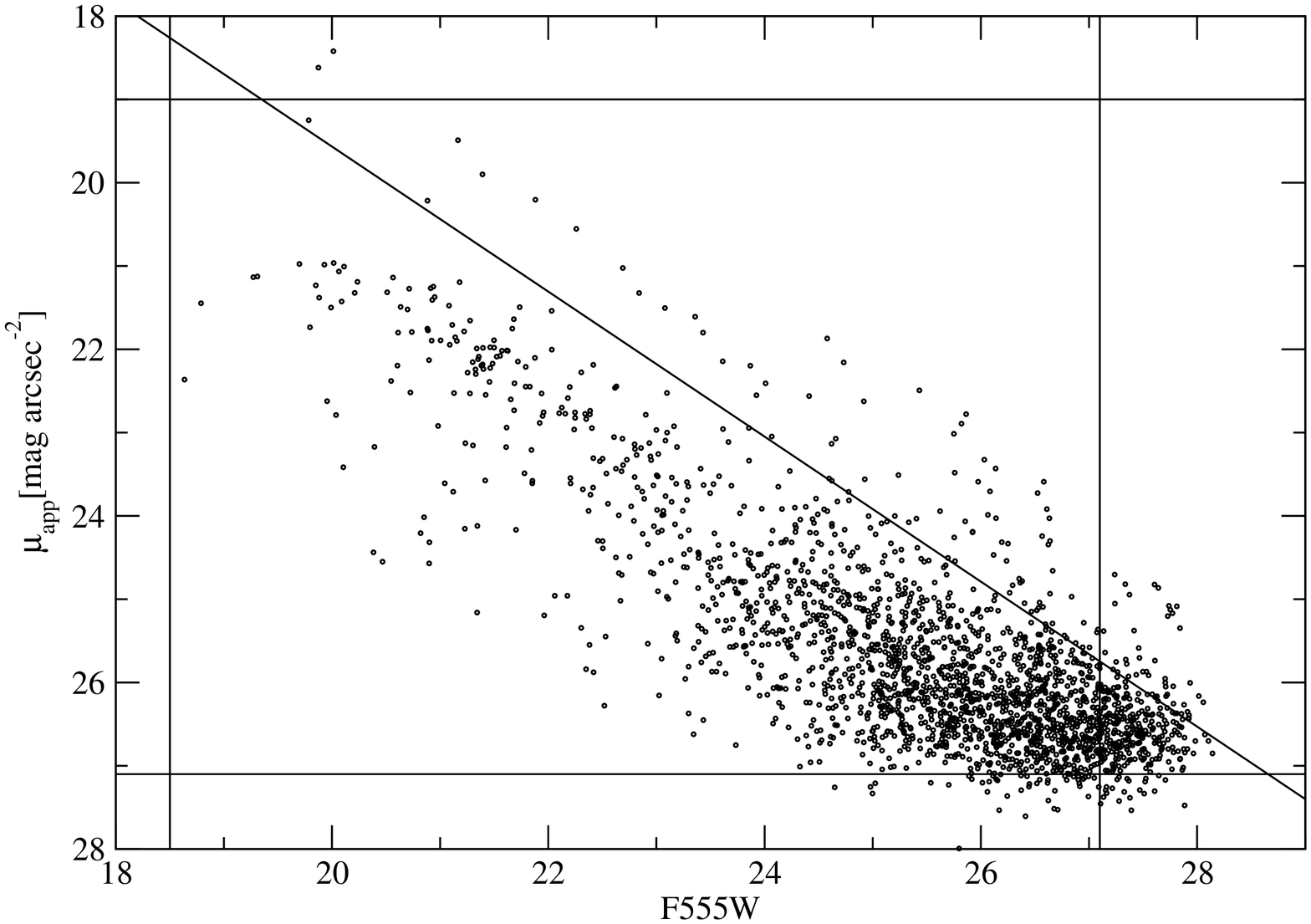} & \includegraphics[width=6cm,angle=270]{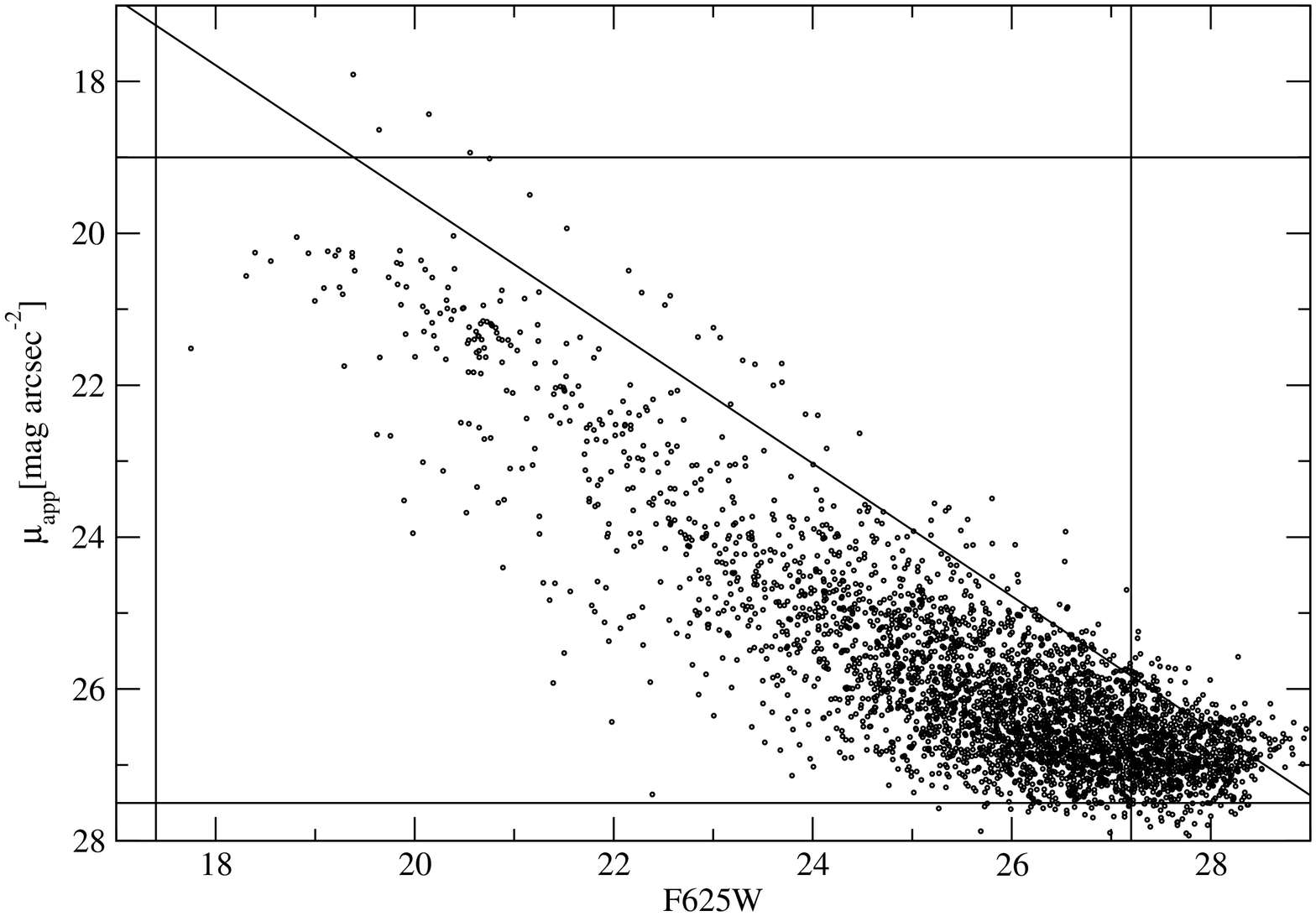} \\
\includegraphics[width=6cm,angle=270]{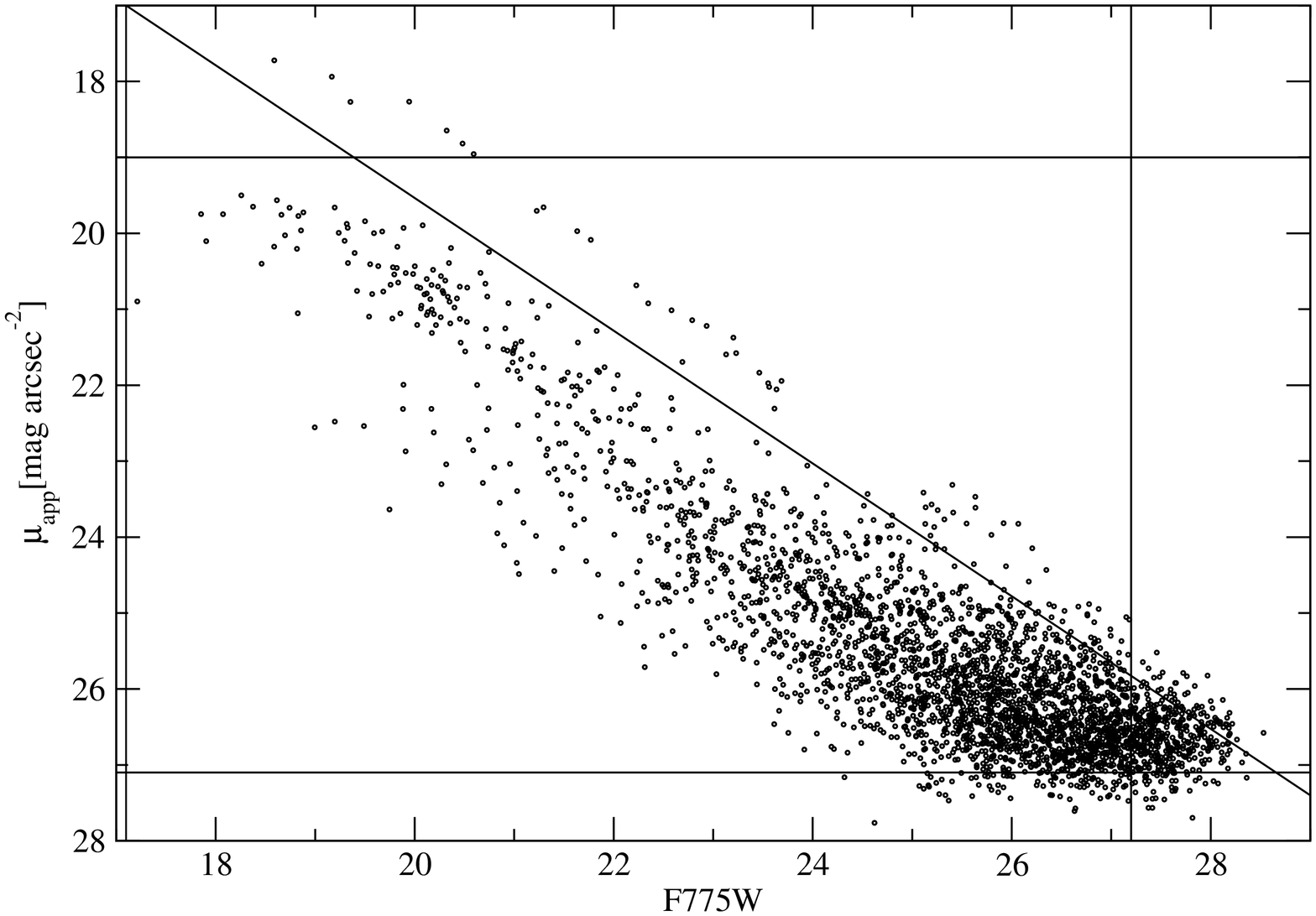} & \includegraphics[width=6cm,angle=270]{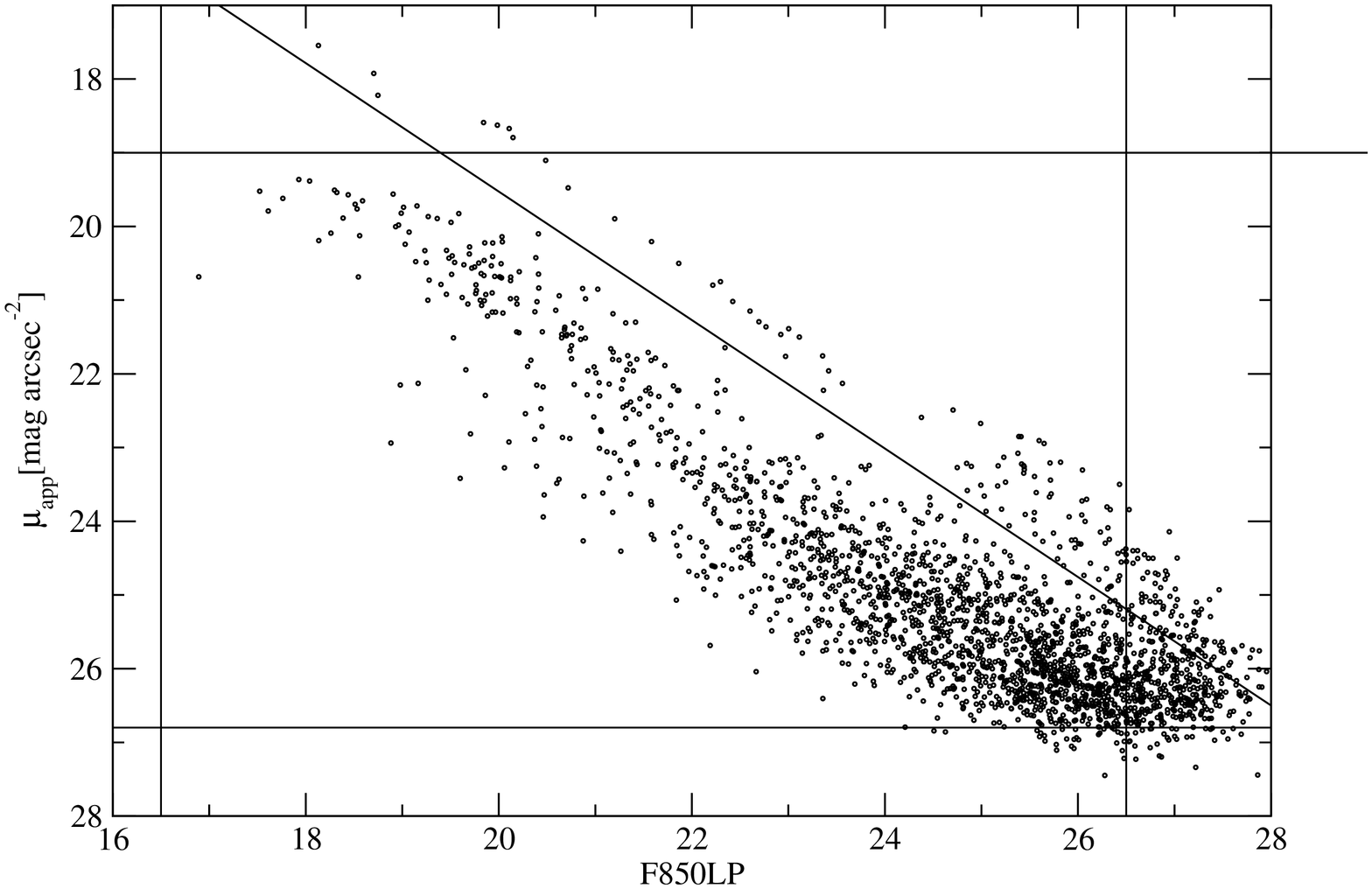} \\
\end{tabular}
\caption{ Mean central surface brightness vs. total magnitudes for objects in the
Abell 1703 field. Selection lines show the star-galaxy separation, the surface
brightness threshold and the magnitude completeness limit. Similar figures for all 
the other objects in our sample can be found in the electronic version of this 
paper.}
\end{figure*}

\clearpage

\begin{figure*}
\begin{tabular}{cc}
\includegraphics[width=6cm,angle=270]{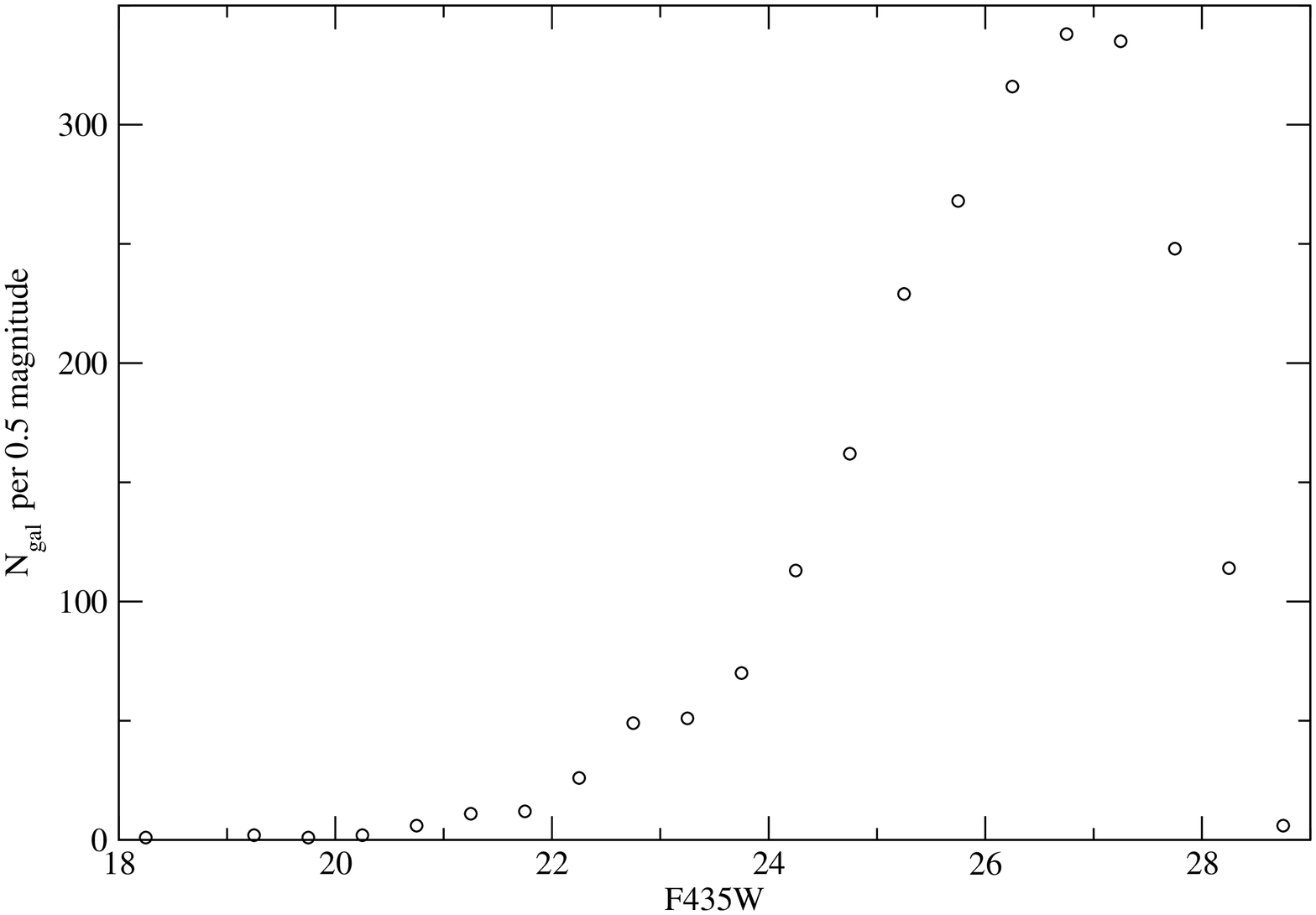} & \includegraphics[width=6cm,angle=270]{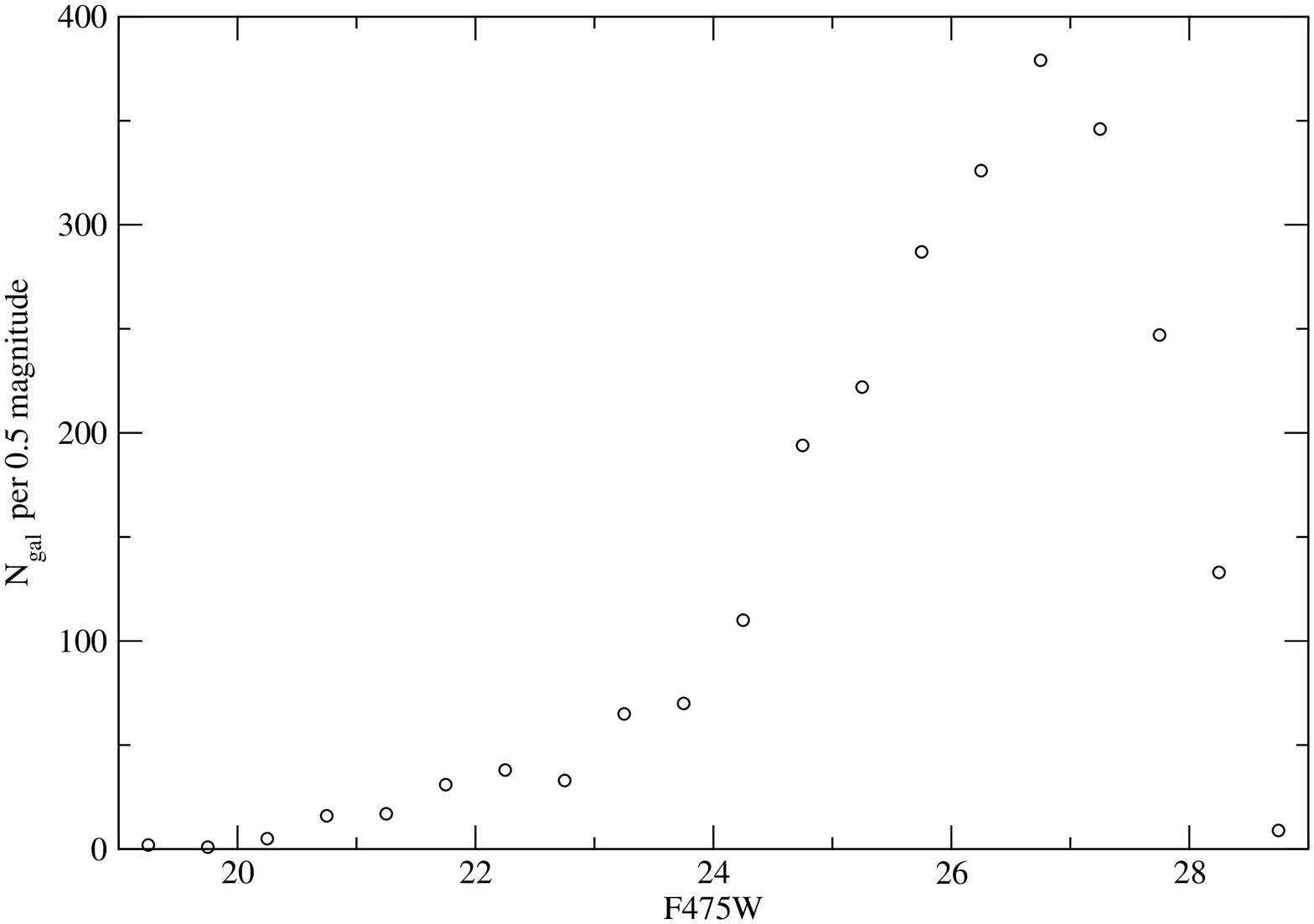} \\
\includegraphics[width=6cm,angle=270]{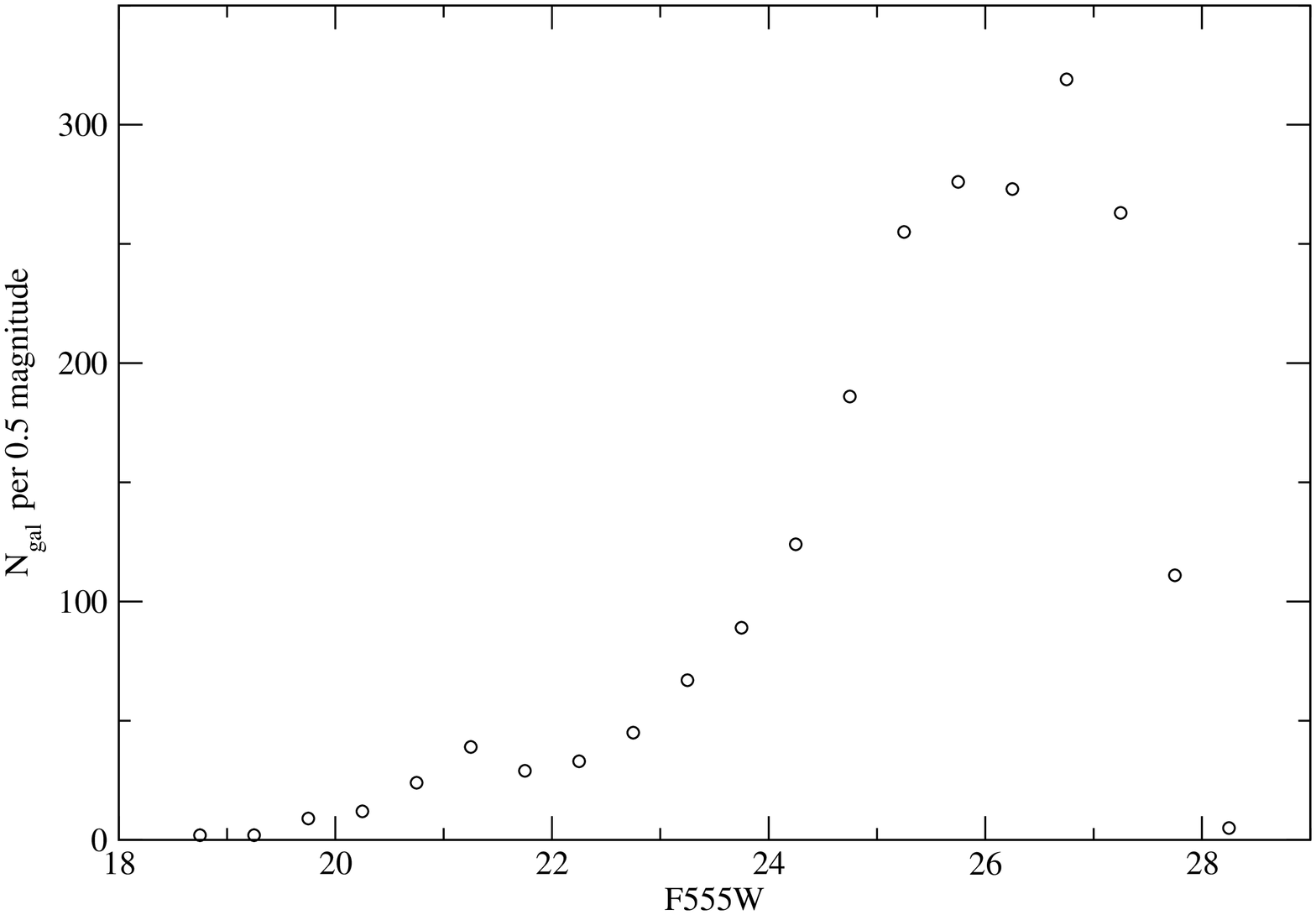} & \includegraphics[width=6cm,angle=270]{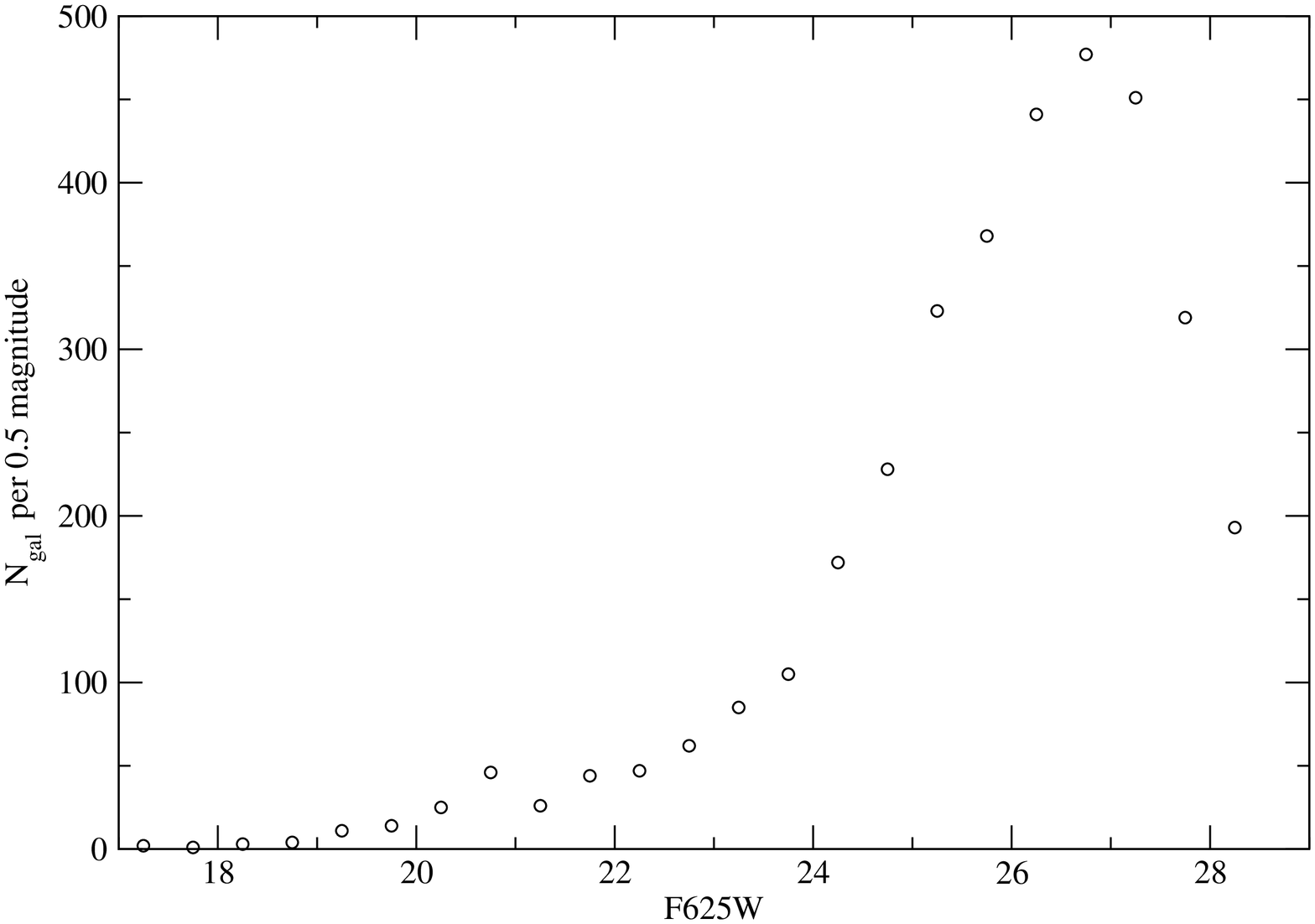} \\
\includegraphics[width=6cm,angle=270]{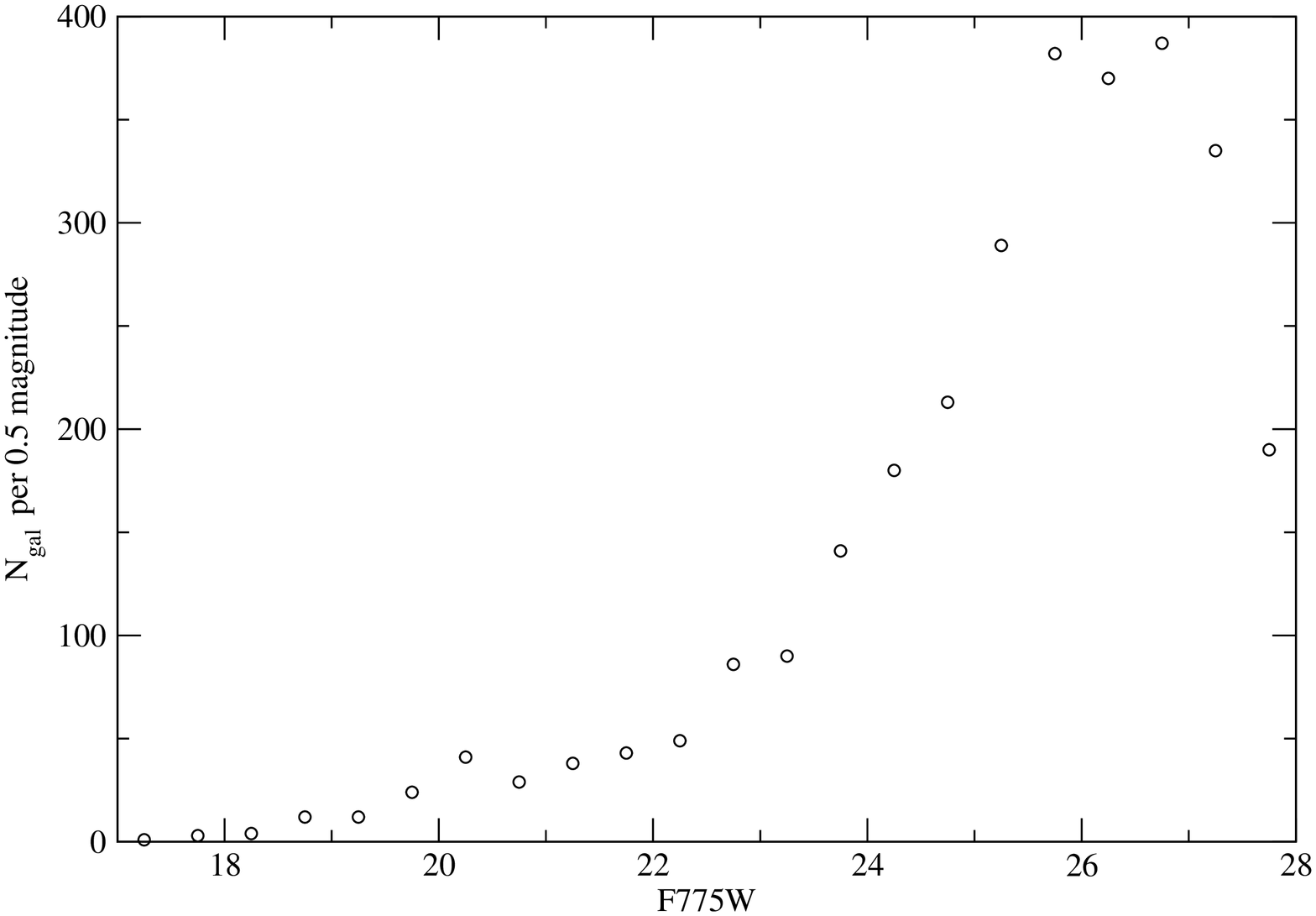} & \includegraphics[width=6cm,angle=270]{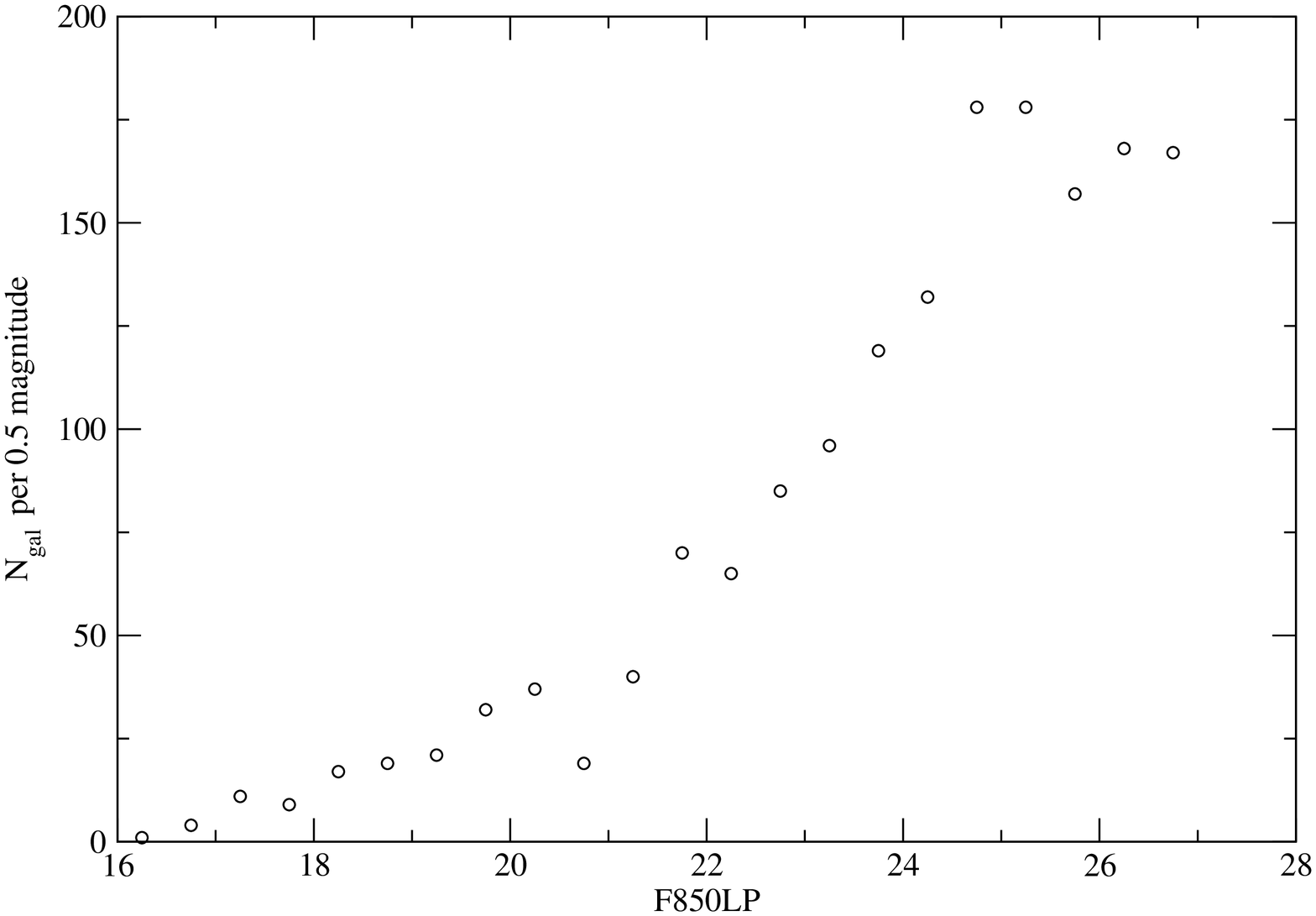} \\
\end{tabular}
\caption{Raw galaxy number counts in the Abell 1703 field. See the electronic
version of this paper for the relative figures for all other targets.}
\end{figure*}
\clearpage

\begin{figure*}
\begin{tabular}{cc}
\includegraphics[width=5cm,angle=270]{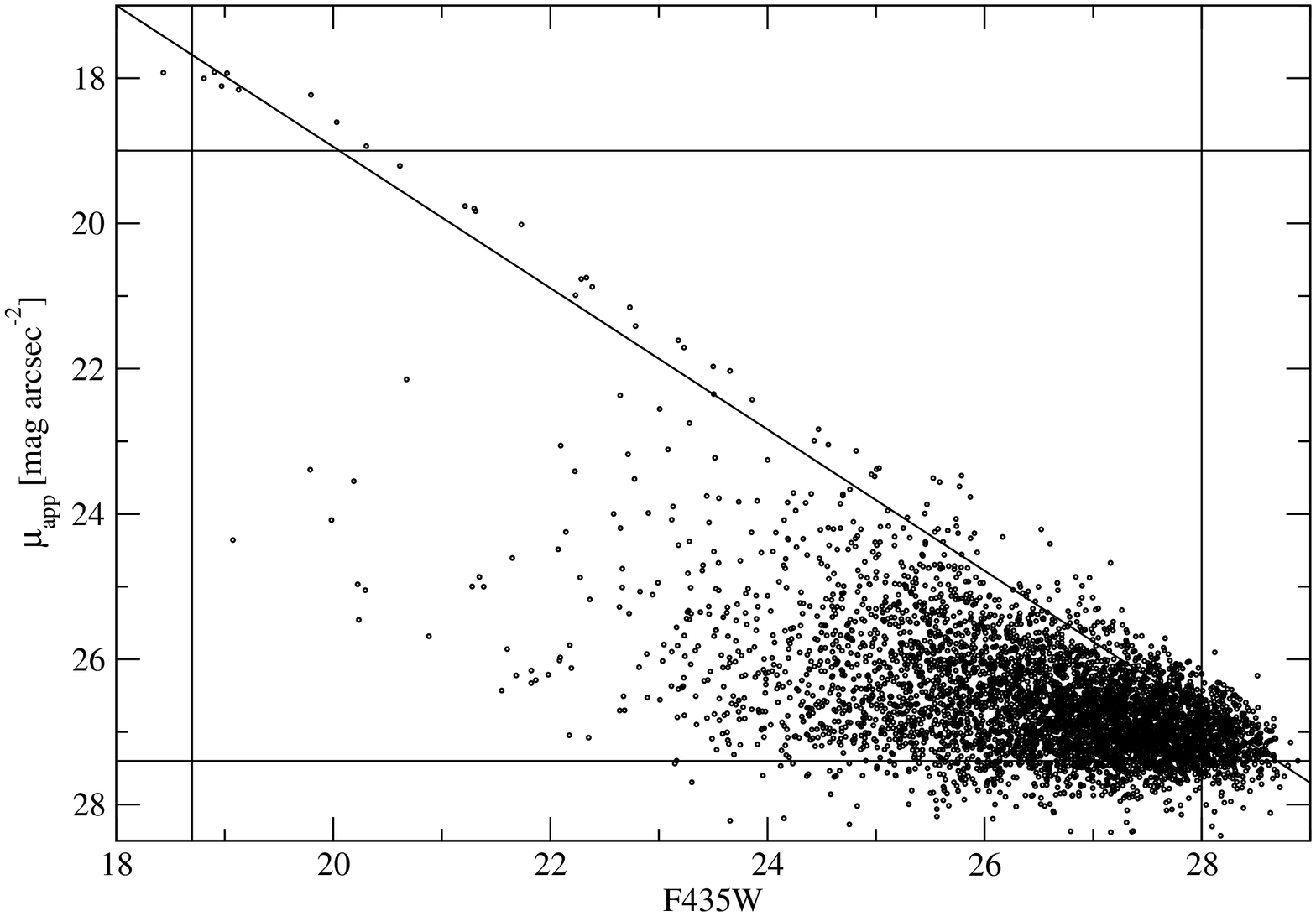} & \includegraphics[width=5cm,angle=270]{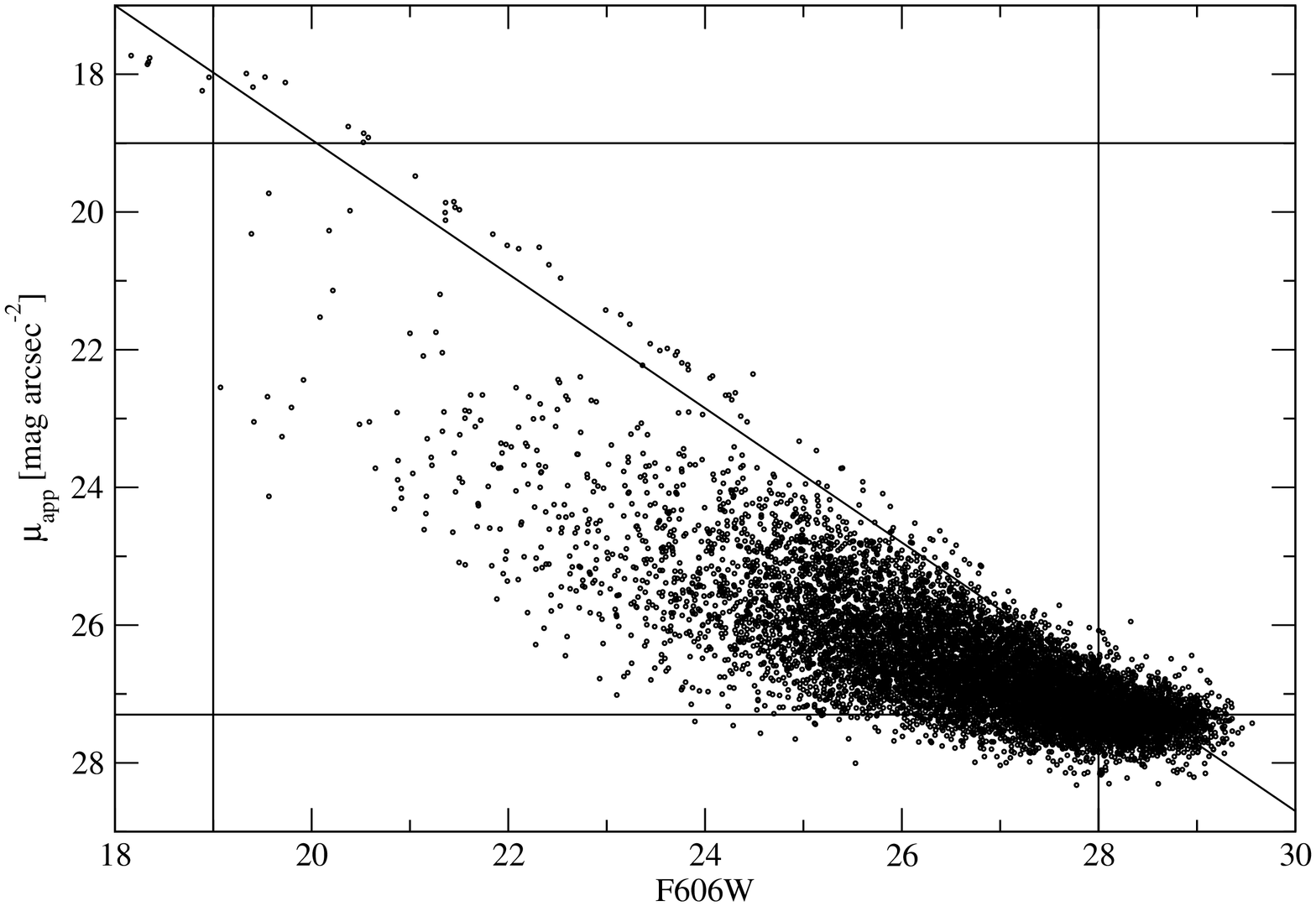} \\
\includegraphics[width=5cm,angle=270]{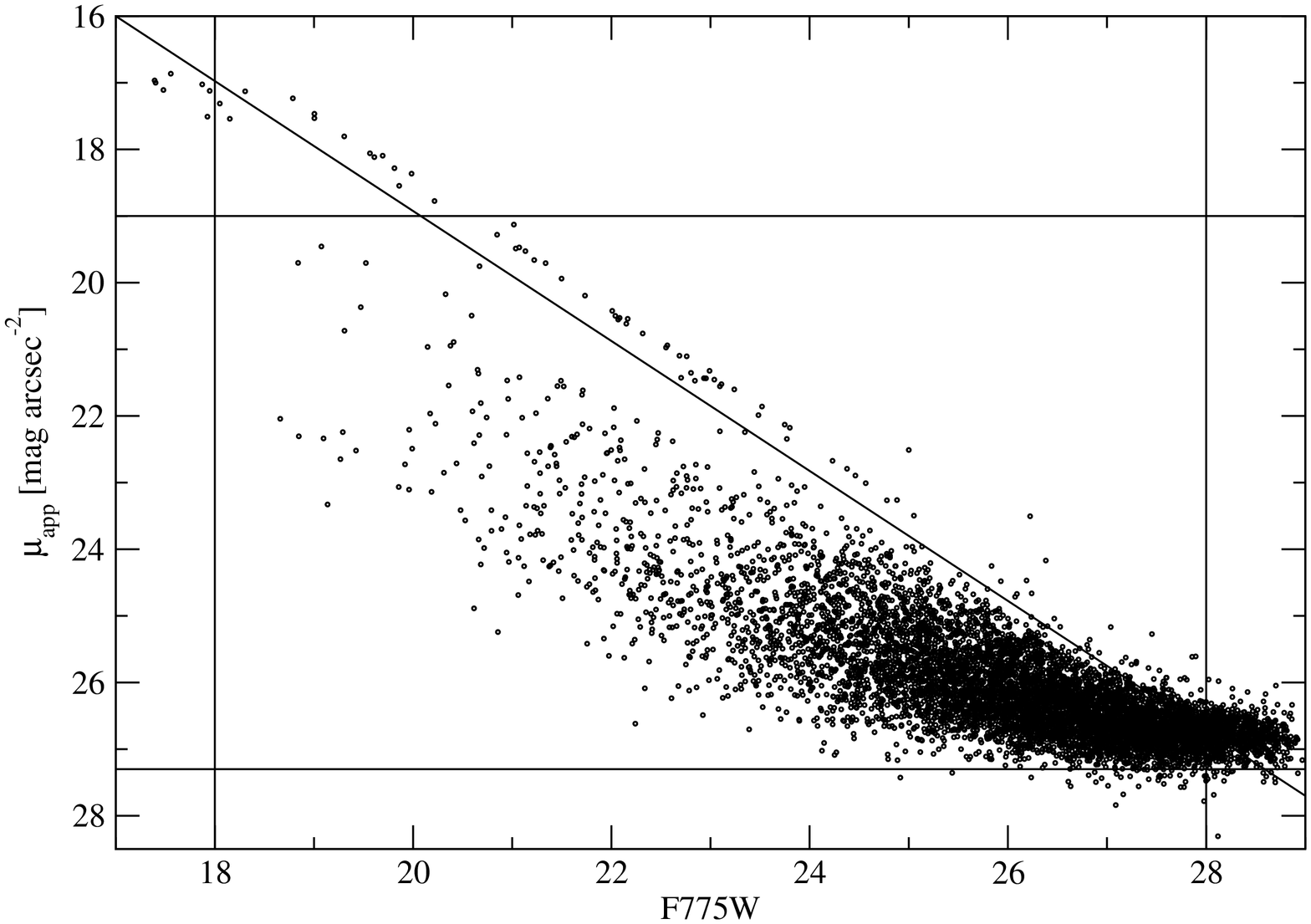} & \includegraphics[width=5cm,angle=270]{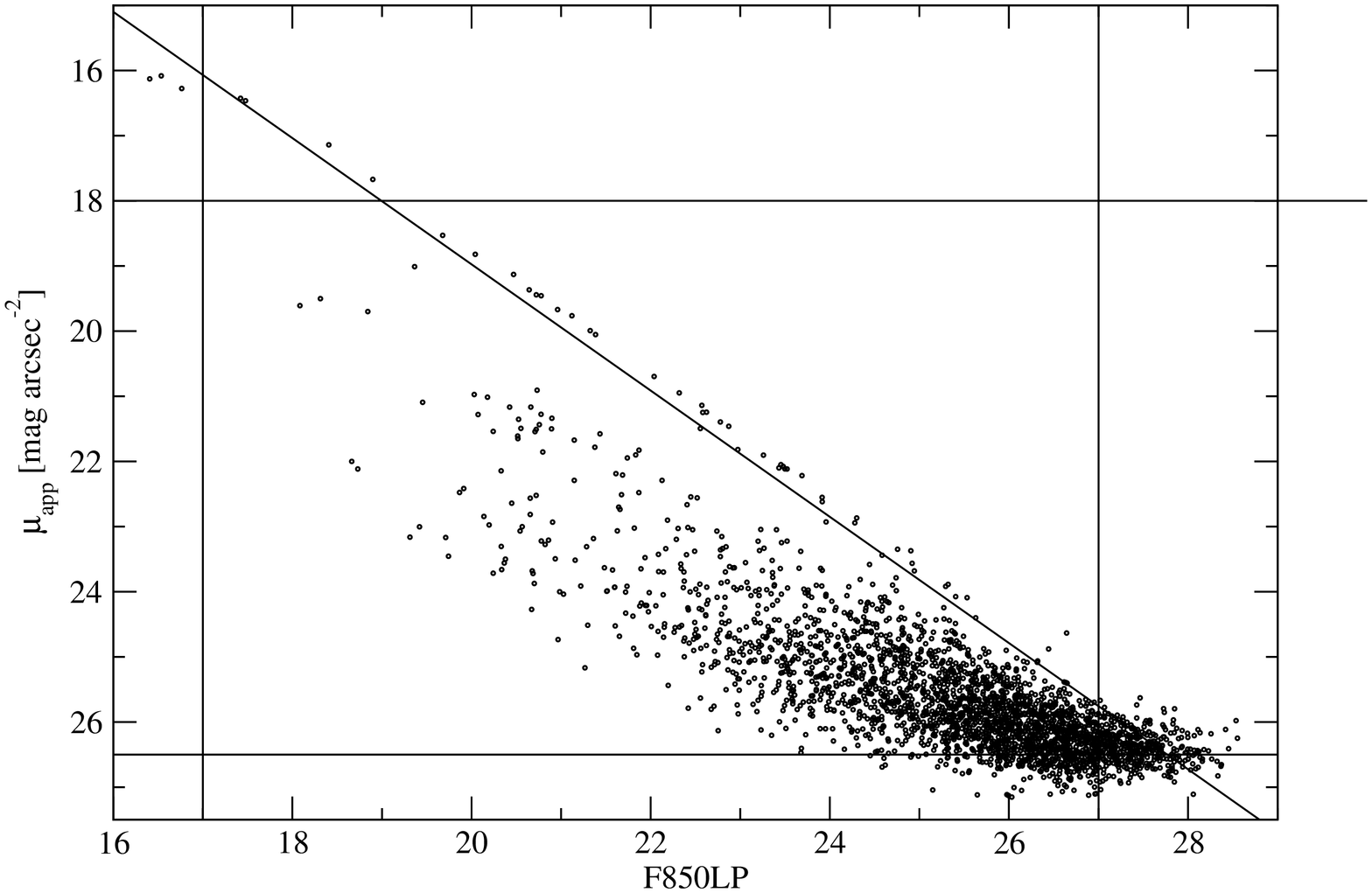} \\
\end{tabular}
\caption{Same as Fig.~2, but for the combined GOODS fields. We only plot a random
selection of 10\% of the objects in the GOODS fields, for clarity.}
\end{figure*}
\clearpage

\begin{figure*}
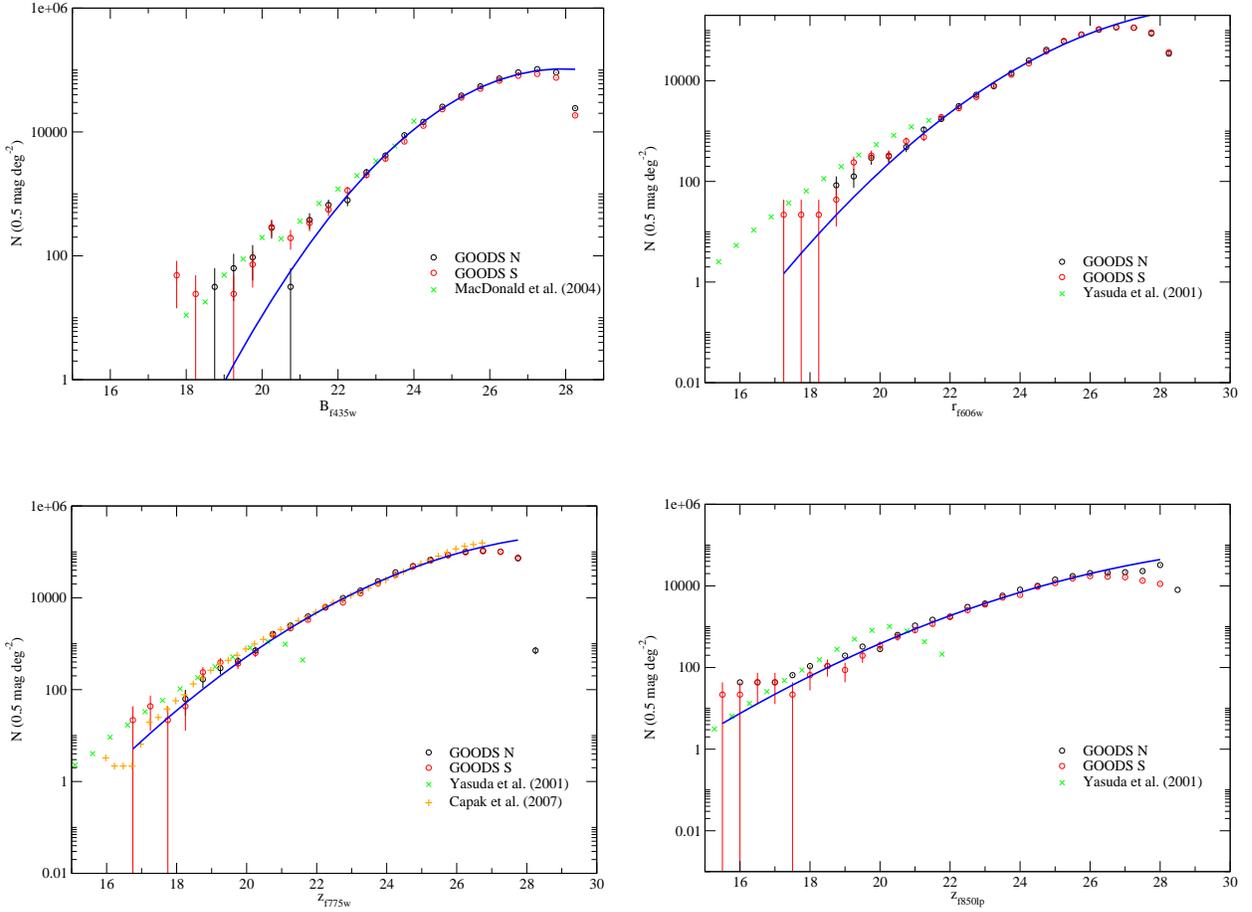

\begin{tabular}{cc}
\includegraphics[width=8cm]{f5a.eps} & \includegraphics[width=8cm]{f5b.eps} \\
 &  \\
\includegraphics[width=8cm]{f5c.eps} & \includegraphics[width=8cm]{f5d.eps} \\
\end{tabular}
\caption{Number counts for the GOODS fields and quadratic fits. Literature
counts (see text) are also shown.}
\end{figure*}
\clearpage

\begin{figure*}
\plottwo{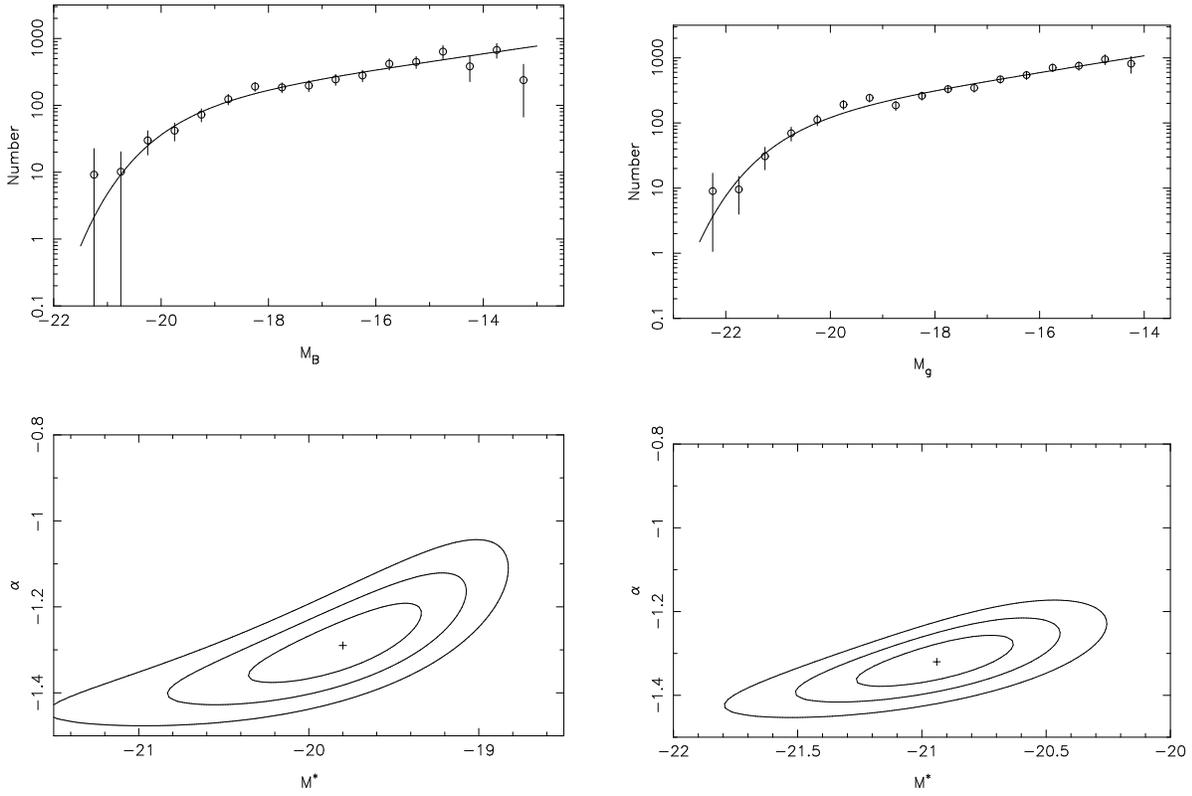}{f6b.ps} \\
\caption{Composite luminosity functions, best fits and error ellipses for the $B$ and $g$ bands}
\end{figure*}
\clearpage

\begin{figure*}
\setcounter{figure}{5}
\plottwo{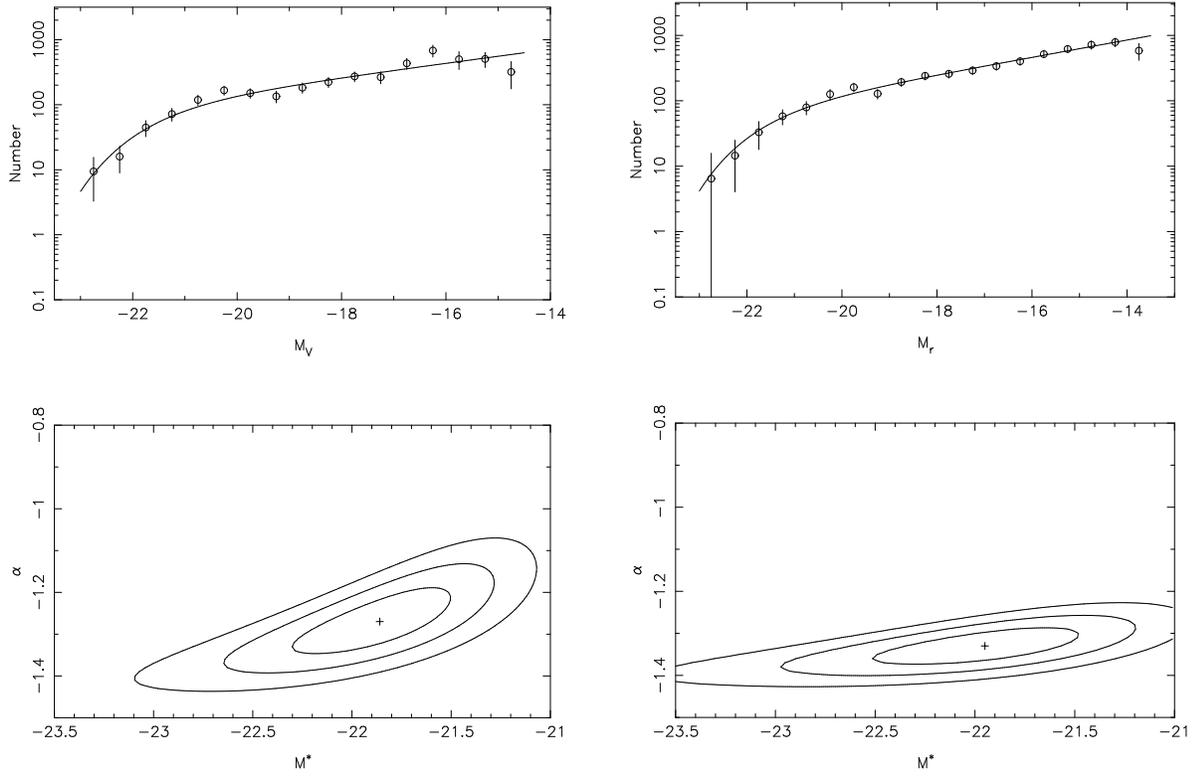}{f6d.ps} \\
\caption{(Continued): for the $V$ and $r$ bands}
\end{figure*}
\clearpage

\begin{figure*}
\setcounter{figure}{5}
\plottwo{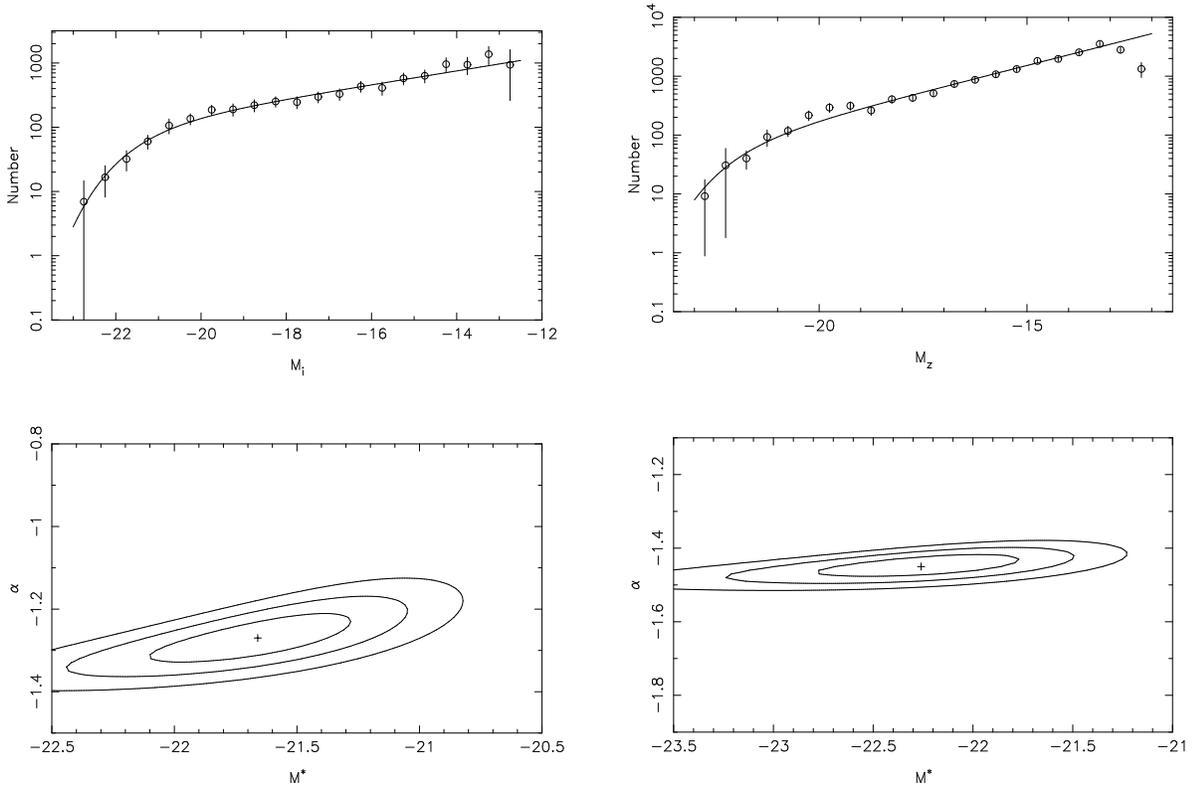}{f6f.ps} \\
\caption{(Continued): for the $i$ and $z$ bands}
\end{figure*}
\clearpage

\begin{figure}
\includegraphics[width=10cm,angle=270]{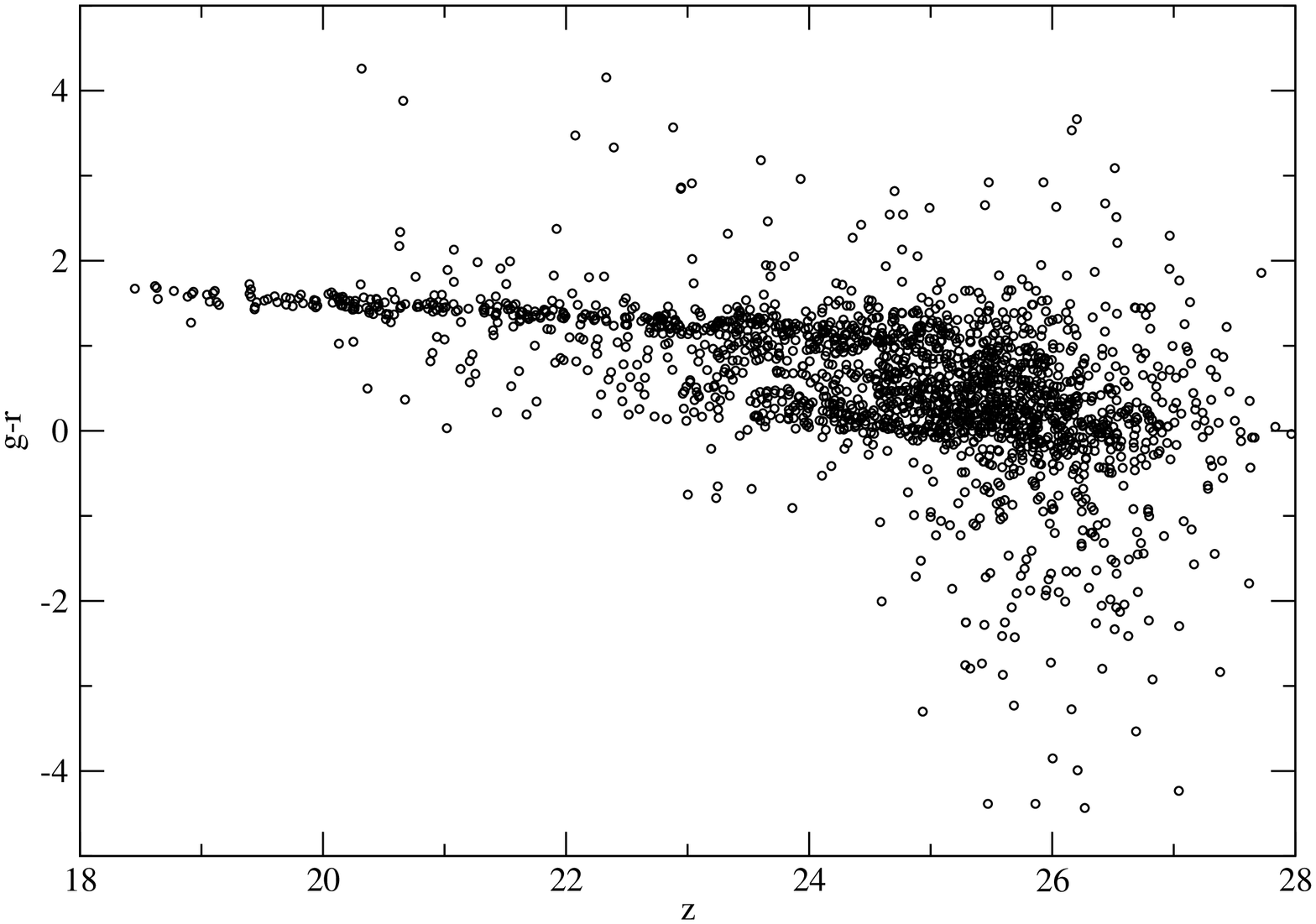}
\caption{The color magnitude diagram for galaxies in A1703}
\end{figure}

\clearpage

\begin{deluxetable}{lcccccccl}
\tabletypesize{\scriptsize}
\rotate
\tablecaption{Archival HST cluster observations}
\tablewidth{0pt}
\tablehead{
\colhead{Cluster} & \colhead{redshift} & \colhead{$B$ Exposure} & \colhead{$g$} & \colhead{$V$} &
\colhead{$r$} & \colhead{$i$} & \colhead{$z$} & \colhead{Proposal ID}
}
\startdata
Abell 1413 & 0.143 & \nodata & \nodata & \nodata & \nodata & 2440 & 2560 & 9292 \\
Abell 2218 & 0.171 &  7048 & 5640 & 7048 & 8386 & 10732 & 5640 & 9452, 9717, 10325 \\
Abell 1689 & 0.183 & \nodata & 9500 & \nodata & 9500 & 11800 & 16600 & 9289 \\
Abell 1703 & 0.258 & 7050 & 5564 & 5564 & 9834 & 11128 & 17800 & 10325 \\
MS 1358.4+6245 & 0.328 & 7928 & 5470 & 5482 & 9196 & 10952 & 15017 & 9292, 9717, 10325 \\
Cl 0024.0+1652 & 0.395 & 6435 & 5072 & 5072 & 8971 & 10144 & 16328 & 10325 \\
\enddata

\end{deluxetable}


\clearpage

\begin{deluxetable}{lcccccc}
\tabletypesize{\scriptsize}
\rotate
\tablecaption{Cluster properties}
\tablewidth{0pt}
\tablehead{
\colhead{Cluster} & \colhead{redshift} & \colhead{$\sigma$ [km s$^{-1}$]} & \colhead{$T$ (KeV)} 
& \colhead{$L_X$ ($10^{44}$ [ergs s$^{-1}$]} & \colhead{$r_{200}$ (Mpc $h^{-1}$)} & \colhead{Coverage (\% $r_{200}$)} 
}
\startdata
Abell 1413 & 0.143 & \nodata & $8.85 \pm 0.50$ & $36.01 \pm 4.54$ & 1.71 & 21 \\
Abell 2218 & 0.171 & $1800 \pm 207$ & $9.02^{+0.40}_{-0.30}$ & $55.73 \pm 8.92$ & 2.94 & 14 \\
Abell 1689 & 0.183 & $1370^{+180}_{-220}$ & $7.10 \pm 0.20$ & $21.96 \pm 2.01$ & 2.24 & 19 \\
Abell 1703 & 0.258 & \nodata & \nodata & \nodata & \nodata & \nodata \\
MS 1358.4+6245 & 0.328 & $937 \pm 54$ & $7.50 \pm 4.30$ & $21.81 \pm 3.81$ & 2.40 & 28 \\
Cl 0024.0+1652 & 0.395 & $1339 \pm 233$ & \nodata & $4.26 \pm 0.01$ & 1.95 & 39 \\
\enddata

\end{deluxetable}


\clearpage


\clearpage

\begin{deluxetable}{rr}
\tablewidth{0pt}
\tablecaption{Sextractor parameters}
\tablehead{
\colhead{Parameter}           & \colhead{Value}}
\startdata
DETECT MINAREA & 7.0 pixels \\
DETECT THRESH & 2.0 \\
ANALYSIS THRESH & 2.0 \\
DEBLEND NTHRESH & 32 \\
PHOT APERTURES & 1.5 \\
PHOT AUTOPARAMS & 2.5, 3.5 \\
\enddata

\end{deluxetable}

\clearpage

\begin{deluxetable}{cccc}
\tablewidth{0pt}
\tablecaption{Quadratic fits to GOODS counts}
\tablehead{
\colhead{Band} & \colhead{a0} & \colhead{a1} & \colhead{a2}}
\startdata
$g$ (F475W) & $-44.46$ & 3.544 & $-0.06346$ \\
$r$ (F606W) & $-23.58$ & 1.928 & $-0.03196$ \\
$i$ (F775W) & $-18.51$ & 1.590 & $-0.02643$ \\
$z$ (F850LP) & $-10.50$ & 0.9377 & $-0.01417$ \\
\enddata

\end{deluxetable}

\clearpage

\begin{deluxetable}{cccc}
\tablewidth{0pt}
\tablecaption{Best fit parameters for composite LFs}
\tablehead{
\colhead{Band} & \colhead{$M^*$} & \colhead{$\alpha$} & \colhead{$\chi^2_\nu$}}
\startdata
$B$ (F435W) & $-19.80 \pm 0.27$ & $-1.29 \pm 0.06$ & $ 0.59 $ \\
$g$ (F475W) & $-20.94 \pm 0.17$ & $-1.31 \pm 0.04$ & $ 0.95 $ \\
$V$ (F555W) & $-21.86 \pm 0.27$ & $-1.27 \pm 0.05$ & $ 1.08 $ \\
$r$ (F625W) & $-21.95 \pm 0.29$ & $-1.33 \pm 0.03$ & $ 0.43 $ \\
$i$ (F775W) & $-21.66 \pm 0.27$ & $-1.27 \pm 0.04$ & $ 0.37 $ \\
$z$ (F850LP) & $-22.26 \pm 0.30$ & $-1.45 \pm 0.02$ & $ 0.94$ \\
\enddata

\end{deluxetable}

\end{document}